\documentclass[twocolumn,showpacs,aps,prl,superscriptaddress]{revtex4}

\usepackage{graphicx}
\usepackage{dcolumn}
\usepackage{amsmath}
\usepackage{epsfig,color}

\RequirePackage{xspace}

\usepackage{relsize}
\def\babar{\mbox{\slshape B\kern-0.1em{\smaller A}\kern-0.1em
    B\kern-0.1em{\smaller A\kern-0.2em R}}}

\def\epem       {\ensuremath{e^+e^-}\xspace}

\def\mumu       {\ensuremath{\mu^+\mu^-}\xspace}

\def\qqbar {\ensuremath{q\overline q}\xspace}

\def\ubar  {\ensuremath{\overline u}\xspace}

\def\dbar  {\ensuremath{\overline d}\xspace}

\def\cbar  {\ensuremath{\overline c}\xspace}
\def\ccbar {\ensuremath{c\overline c}\xspace}

\def\pip   {\ensuremath{\pi^+}\xspace}
\def\pim   {\ensuremath{\pi^-}\xspace}

\def\Kbar  {\kern 0.2em\overline{\kern -0.2em K}{}\xspace}

\def\Kz    {\ensuremath{K^0}\xspace}
\def\Kzb   {\ensuremath{\Kbar^0}\xspace}
\def\KzKzb {\ensuremath{\Kz \kern -0.16em \Kzb}\xspace}
\def\Kp    {\ensuremath{K^+}\xspace}
\def\Km    {\ensuremath{K^-}\xspace}

\def\KpKm  {\ensuremath{\Kp \kern -0.16em \Km}\xspace}
\def\KS    {\ensuremath{K^0_{\scriptscriptstyle S}}\xspace}

\def\Dbar    {\kern 0.2em\overline{\kern -0.2em D}{}\xspace}

\def\Dz      {\ensuremath{D^0}\xspace}
\def\Dzb     {\ensuremath{\Dbar^0}\xspace}
\def\DzDzb   {\ensuremath{\Dz {\kern -0.16em \Dzb}}\xspace}
\def\Dp      {\ensuremath{D^+}\xspace}
\def\Dm      {\ensuremath{D^-}\xspace}

\def\DpDm    {\ensuremath{\Dp {\kern -0.16em \Dm}}\xspace}

\def\Dstarzb {\ensuremath{\Dbar^{*0}}\xspace}

\def\Bbar    {\kern 0.18em\overline{\kern -0.18em B}{}\xspace}

\def\BB      {\ensuremath{B\Bbar}\xspace} 
\def\Bz      {\ensuremath{B^0}\xspace}
\def\Bzb     {\ensuremath{\Bbar^0}\xspace}
\def\BzBzb   {\ensuremath{\Bz {\kern -0.16em \Bzb}}\xspace}
\def\Bu      {\ensuremath{B^+}\xspace}
\def\Bub     {\ensuremath{B^-}\xspace}

\def\Bm      {\ensuremath{\Bub}\xspace}

\def\BpBm    {\ensuremath{\Bu {\kern -0.16em \Bub}}\xspace}

\def\BorBbar    {\kern 0.18em\optbar{\kern -0.18em B}{}\xspace}
\def\DorDbar    {\kern 0.18em\optbar{\kern -0.18em D}{}\xspace}
\def\KorKbar    {\kern 0.18em\optbar{\kern -0.18em K}{}\xspace}

\def\jpsi     {\ensuremath{{J\mskip -3mu/\mskip -2mu\psi\mskip 2mu}}\xspace}
\def\psitwos  {\ensuremath{\psi{(2S)}}\xspace}

\mathchardef\Upsilon="7107
\def\Y#1S{\ensuremath{\Upsilon{(#1S)}}\xspace}

\def\FourS {\Y4S}

\mathchardef\Deltares="7101
\mathchardef\Xi="7104
\mathchardef\Lambda="7103
\mathchardef\Sigma="7106
\mathchardef\Omega="710A

\def\Deltabar{\kern 0.25em\overline{\kern -0.25em \Deltares}{}\xspace}
\def\Lbar{\kern 0.2em\overline{\kern -0.2em\Lambda\kern 0.05em}\kern-0.05em{}\xspace}
\def\Sigbar{\kern 0.2em\overline{\kern -0.2em \Sigma}{}\xspace}
\def\Xibar{\kern 0.2em\overline{\kern -0.2em \Xi}{}\xspace}
\def\Obar{\kern 0.2em\overline{\kern -0.2em \Omega}{}\xspace}
\def\Nbar{\kern 0.2em\overline{\kern -0.2em N}{}\xspace}
\def\Xb{\kern 0.2em\overline{\kern -0.2em X}{}\xspace}

\def\BR         {{\ensuremath{\cal B}\xspace}}

\def\mes        {\mbox{$m_{\rm ES}$}\xspace}

\def\DeltaE     {\mbox{$\Delta E$}\xspace}

\newcommand{\tev}{\ensuremath{\mathrm{\,Te\kern -0.1em V}}\xspace}
\newcommand{\gev}{\ensuremath{\mathrm{\,Ge\kern -0.1em V}}\xspace}
\newcommand{\mev}{\ensuremath{\mathrm{\,Me\kern -0.1em V}}\xspace}
\newcommand{\kev}{\ensuremath{\mathrm{\,ke\kern -0.1em V}}\xspace}
\newcommand{\ev}{\ensuremath{\mathrm{\,e\kern -0.1em V}}\xspace}
\newcommand{\gevc}{\ensuremath{{\mathrm{\,Ge\kern -0.1em V\!/}c}}\xspace}
\newcommand{\mevc}{\ensuremath{{\mathrm{\,Me\kern -0.1em V\!/}c}}\xspace}
\newcommand{\gevcc}{\ensuremath{{\mathrm{\,Ge\kern -0.1em V\!/}c^2}}\xspace}
\newcommand{\mevcc}{\ensuremath{{\mathrm{\,Me\kern -0.1em V\!/}c^2}}\xspace}

\def\invfb   {\ensuremath{\mbox{\,fb}^{-1}}\xspace}

\def\mus  {\ensuremath{\rm \,\mus}\xspace}

\def\mus        {\ensuremath{\,\mu{\rm s}}\xspace}    

\def\to                 {\ensuremath{\rightarrow}\xspace}

\def\pep2{PEP-II}

\def\gsim{{~\raise.15em\hbox{$>$}\kern-.85em
          \lower.35em\hbox{$\sim$}~}\xspace}
\def\lsim{{~\raise.15em\hbox{$<$}\kern-.85em
          \lower.35em\hbox{$\sim$}~}\xspace}

\xspace

\def\jetset74   {\mbox{\tt Jetset \hspace{-0.5em}7.\hspace{-0.2em}4}\xspace}

\newcommand{\BABARPubYear}    {05}
\newcommand{\BABARPubNumber}  {038}

\newcommand{\SLACPubNumber} {11370}

\def\figurebox#1#2#3{
    \def\arg{#3}
    \ifx\arg\empty
    {\hfill\vbox{\hsize#2\hrule\hbox to #2{\vrule\hfill\vbox to #1{\hsize#2\vfill}\vrule}\hrule}\hfill}
    \else
    {\hfill\epsfbox{#3}\hfill}
    \fi}

\begin{document}

\def\Xz {\ensuremath{X(3872)}\xspace}

\begin{flushleft}
\babar-PUB-\BABARPubYear/\BABARPubNumber\\
SLAC-PUB-\SLACPubNumber\\
\end{flushleft}

\title{\large\boldmath
Study of the $X(3872)$ and $Y(4260)$ in
$B^0\rightarrow J/\psi\pi^+\pi^-K^0$\\ and $B^-\rightarrow J/\psi\pi^+\pi^-K^-$
decays}
\author{B.~Aubert}
\author{R.~Barate}
\author{D.~Boutigny}
\author{F.~Couderc}
\author{Y.~Karyotakis}
\author{J.~P.~Lees}
\author{V.~Poireau}
\author{V.~Tisserand}
\author{A.~Zghiche}
\affiliation{Laboratoire de Physique des Particules, F-74941 Annecy-le-Vieux, France }
\author{E.~Grauges}
\affiliation{IFAE, Universitat Autonoma de Barcelona, E-08193 Bellaterra, Barcelona, Spain }
% ...
\collaboration{The \babar\ Collaboration}
\noaffiliation
\date{\today}

\begin{abstract}
We present results of a search for the $X(3872)$ in $B^0\rightarrow
X(3872)K^0_{\scriptscriptstyle S}, X(3872)\rightarrow J/\psi
\pi^+\pi^-$, improved measurements of $B^-\rightarrow X(3872) K^-$,
and a study of the $J/\psi\pi^+\pi^-$ mass region above the
$X(3872)$. We use 232 million $B \bar B$ pairs collected at the
$\Upsilon(4S)$ resonance with the \mbox{\slshape B\kern-0.1em{\smaller
A}\kern-0.1em B\kern-0.1em {\smaller A\kern-0.2em R}} detector at the
PEP-II $e^+ e^-$ asymmetric-energy storage rings. The results include
the 90\% confidence interval $1.34\times 10^{-6}
<{\cal B}(B^0\rightarrow X(3872)K^0,X\rightarrow J/\psi\pi^+\pi^-) <10.3\times 10^{-6}$ and the
branching fraction
${\cal B}(B^-\rightarrow X(3872)K^-,X\rightarrow J/\psi\pi^+\pi^-)=(10.1\pm2.5\pm1.0)\times
10^{-6}$. We observe a $(2.7\pm 1.3 \pm 0.2){\mathrm{\,Me\kern -0.1em V\!/}c^2}$ mass difference of
the $X(3872)$ produced in the two decay modes. Furthermore, we search
for the $Y(4260)$ in $B$ decays and set the $95\%$ C.L. upper limit
$\BR(\Bm\to Y(4260)\Km,Y(4260)\to\jpsi\pip\pim)<2.9\times10^{-5}$.
\end{abstract}

\pacs{13.25.Hw, 14.40.Gx, 12.39.Mk}

\maketitle

The $X(3872)$ was first observed in the charged \hbox{$B$-meson}
decay~\cite{chargeConj} $\Bm\to\Xz\Km$, $\Xz\to\jpsi\pip\pim$ by the
Belle Collaboration~\cite{Choi:2003ue}. It has been confirmed by the
\babar\ Collaboration~\cite{Aubert:2004ns} and observed inclusively in
the same final state by the CDF and D0
collaborations~\cite{CDF-D0}. This narrow-width particle has a mass
very near the $\Dz\Dstarzb$ threshold and decays into final states
containing charmonium (\jpsi). The most plausible
interpretation~\cite{Eichten:2004uh} was a $1^3D_2$ or $1^1D_2$
$c\bar{c}$ state which would be narrow since it would be forbidden to
decay into open charm $D\bar{D}$ states. However, these candidates
should have large radiative transitions into $\chi_c$ states that have
not been observed~\cite{Choi:2003ue}. Recent studies from Belle that
combine angular and kinematic properties of the $\pi^+\pi^-$ mass,
strongly favor a $J^{PC}=1^{++}$ state~\cite{newbelle}. Other
explanations include $2^1P_1$ $\ccbar$ ($1^{+-}$) or $2^3P_1$ $1^{++}$
states that should be narrow, but are predicted to be about
$100\mevcc$ higher than the $X(3872)$ mass and are not expected to
have a large decay rate into $\jpsi\pi\pi$ final
states~\cite{ted}. Hence, the $X(3872)$ appears not to be a simple
quark model $\qqbar$ meson state.

Many explanations have been proposed for the nature of the $X(3872)$.
Recent interpretations include the diquark-antidiquark
model~\cite{Maiani:2004vq} and the S-wave $\Dz\Dstarzb$ molecule
model~\cite{Braaten:2004ai}. The diquark-antidiquark model predicts a
spectrum of $J=0,1,2$ particles and identifies the $X(3872)$ as its
$1^{++}$ member state with the two quark
combinations $X_u=\left[cu\right]\left[\cbar\ubar\right]$ and
$X_d=\left[cd\right]\left[\cbar\dbar\right]$ with a mass difference
$m(X_d)-m(X_u)\approx (7\pm 2)\mevcc$. In addition, these two states
could form mixed states that are produced in both charged and neutral
\hbox{$B$-meson} decays with different masses and rates depending on
the mixing angle. A search for the predicted charged partner of the
\Xz has been addressed in a previous analysis with an upper limit 
that is still consistent with the model~\cite{Aubert:2004zr}. The
$\Dz\Dstarzb$ molecule model interprets the $X(3872)$ as a loosely
bound $\Dz\Dstarzb$ S-wave state that is produced in weak decays of
the \hbox{$B$-meson} into $\Dz\Dstarzb K$.  In this picture, the
S-wave molecule must form a $J^P=1^+$ state.  From factorization,
heavy-quark symmetry, and isospin symmetry, the decay $\Bz\to\Xz\Kz$
is predicted to be suppressed by an order of magnitude relative to
$\Bm\to\Xz\Km$~\cite{Braaten:2004ai}. To investigate these
predictions, we present in this letter a study of the neutral mode
$\Bz\to\Xz\Kz$, $\Xz\to\jpsi\pip\pim$ and we analyze
$\Bm\to\jpsi\pip\pim\Km$ decays with increased statistics to obtain
improved measurements of $X(3872)\to\jpsi\pip\pim$. In addition, we
examine the higher $\jpsi\pip\pim$ invariant mass region to search for
a structure recently observed in initial state radiation (ISR)
events~\cite{Aubert:2005-ISR}.

The data were collected with the $\babar$ detector at the PEP-II
asymmetric-energy $e^{+}e^{-}$ storage rings on the $\FourS$
resonance. The integrated luminosity of the data used in this analysis
is $211\invfb$; this corresponds to the production of
$(232\pm3)\times 10^6$ \BB pairs.  

The \babar\ detector is described in detail elsewhere~\cite{Aubert:2001tu}.
Charged-particle
trajectories are measured by a combination of a five-layer silicon vertex
tracker (SVT) and a \hbox{40-layer} drift chamber (DCH) in a \hbox{1.5-T}
solenoidal magnetic field. 
For charged-particle identification, 
we combine information from a ring-imaging Cherenkov detector (DIRC) 
and energy-loss measurements provided by the SVT and the DCH.
Photons and electrons are detected in a CsI(Tl)
electromagnetic calorimeter (EMC). Penetrating muons are identified by
resistive-plate chambers in the instrumented magnetic flux return (IFR).

Charged pion candidates are required to be detected in at least 12 DCH
layers and have a transverse momentum greater than $100\mevc$. Kaons,
electrons, and muons are separated from pions based on information from the IFR
and DIRC, energy loss in the SVT and DCH $(dE/dx)$, or the ratio of
the candidate EMC energy deposition to its momentum ($E/p$). Photon
candidates are identified with clusters in the EMC with total energy
$>30\mev$ and a shower shape consistent with that expected 
from a photon.

The $\Bz\to\jpsi\pip\pim\KS$ and $\Bm\to\jpsi\pip\pim\Km$ decays are
reconstructed in the following way. Electron candidates and
bremsstrahlung photons satisfying $2.95<m(\epem(\gamma))<3.14\gevcc$
are used to form $\jpsi\to\epem$ candidates.  A pair of muon
candidates within the mass interval $3.06<m(\mumu)<3.14\gevcc$ is
required for a $\jpsi\to\mumu$ candidate. A mass constraint to the
nominal \jpsi mass~\cite{pdg} is imposed in the fit of the lepton
pairs.  We reconstruct $\KS\to\pip\pim$ candidates from pairs of
oppositely charged tracks forming a vertex with a $\chi^2$ probability
larger than $0.1\%$, a flight-length significance $l/\sigma(l)>3$ and
an invariant mass within $15\mevcc$ of the nominal $\KS$
mass~\cite{pdg}. $\Xz$ candidates are formed by combining \jpsi
candidates with two oppositely charged pion candidates fitted to a
common vertex. Finally, we form $\Bz(\Bm)$ candidates by combining
$\Xz$ candidates with $\KS(\Km)$ candidates. To suppress continuum
background, we select only events with a ratio of the second to the zeroth
Fox-Wolfram moment~\cite{Fox1979} less than $0.5$.

We use two kinematic variables to identify signal events from $B$ decays: the
difference between the energy of the $B$ candidate and the beam energy,
$\DeltaE\equiv E_B^*-\sqrt{s}/2$, and the energy-substituted mass
$\mes\equiv\sqrt{(s/2+{\bf p}_i\cdot{\bf p}_B)^2/E_i^2 - {\bf p}_B^2}$. Here
$(E_i,{\bf p}_i)$ is the four-vector (in the laboratory frame) and $\sqrt{s}$ is the
center-of-mass (CM) energy of the $\epem$ system. $E_B^*$ is the energy of the
$B$ candidate in the CM system and ${\bf p}_B$ the momentum in the
laboratory frame. The signature of signal events is $\Delta E\approx 0$, and $\mes
\approx m_{B}$ where $m_{B}$ is the mass of the \hbox{$B$-meson}~\cite{pdg}.

\begin{figure}[ht]
\begin{center}
  \includegraphics[width=0.23\textwidth]{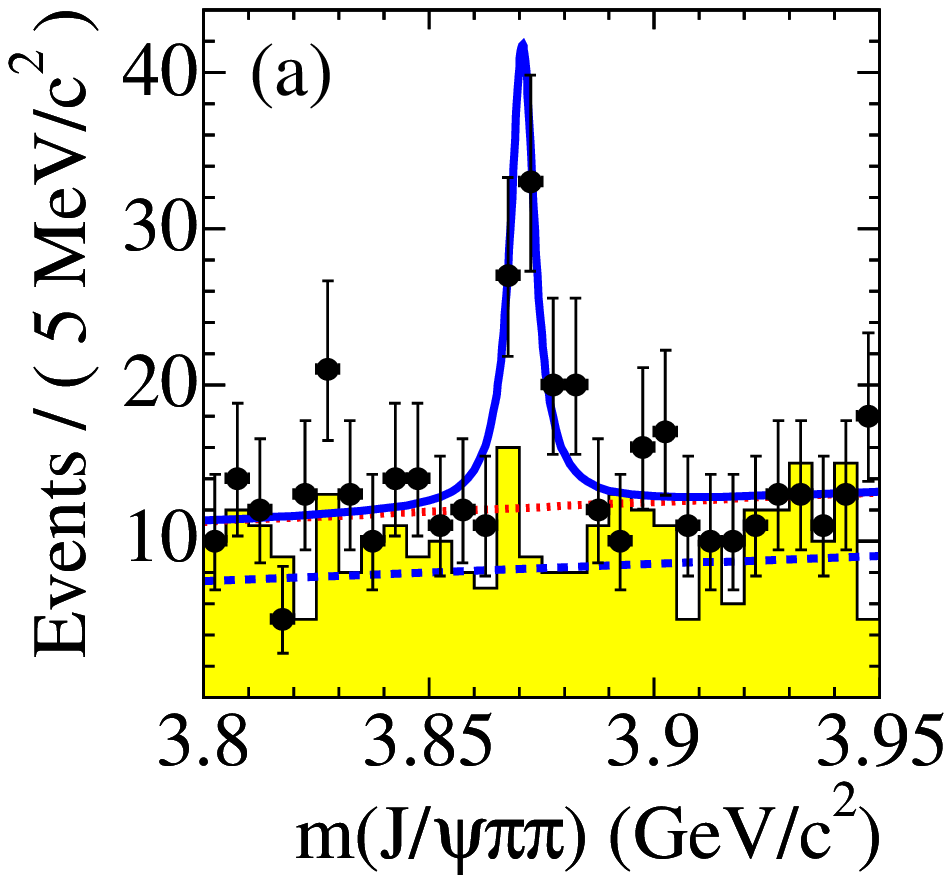}
  \includegraphics[width=0.23\textwidth]{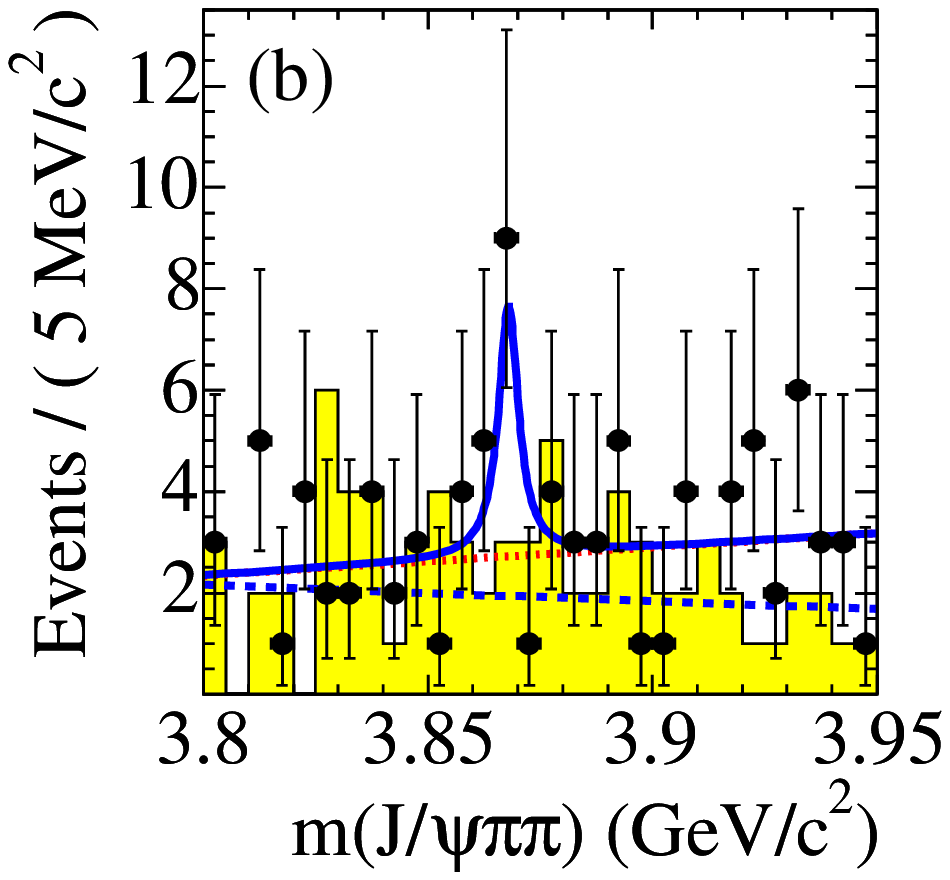}
  \includegraphics[width=0.23\textwidth]{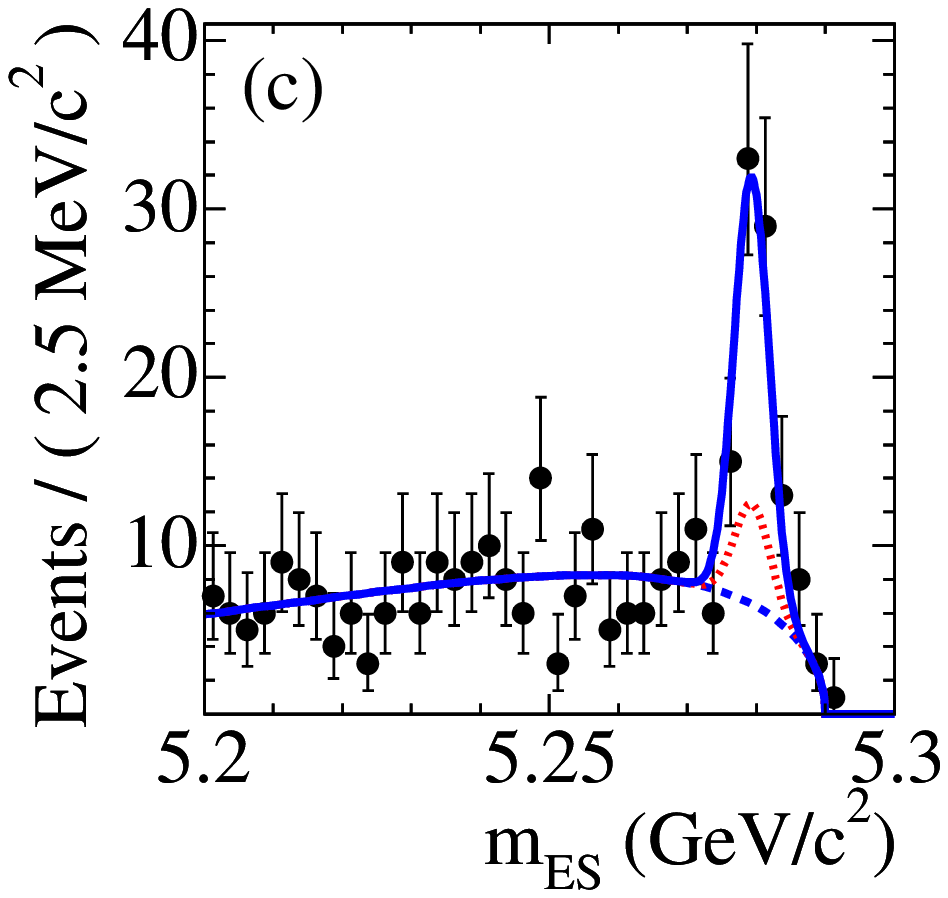}
  \includegraphics[width=0.23\textwidth]{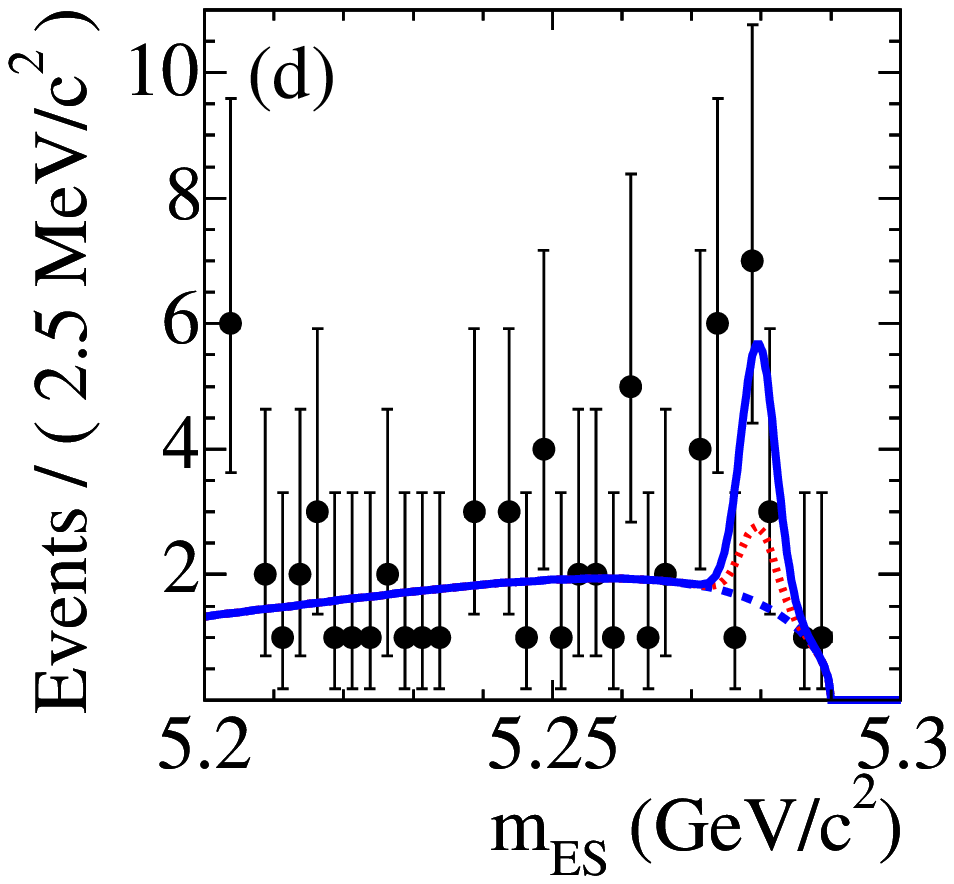}
  \caption{\label{fig:fit}
  Signal region projections of $m(\jpsi\pip\pim)$ and \mes for
  $\Bm\to\Xz\Km$ (a,c) and $\Bz\to\Xz\KS$ (b,d). The dashed line
  represents the combinatorial background PDF, the dotted line
  represents the sum of the combinatorial and peaking background PDF,
  and the solid line the sum of all background plus the signal
  PDF. The shaded area shows events in the \mes sideband region
  $|\mes-5260\mevcc|<6\mevcc$. }
\end{center}
\end{figure}

We optimize the signal selection criteria by maximizing the ratio
$n_s^{mc}/(3/2+\sqrt{n_b^{mc}})$~\cite{Punzi:2003bu} where $n_s^{mc}$
($n_b^{mc}$) are the number of reconstructed Monte Carlo (MC) signal
(background) events.  The optimization was performed by varying the
selection criteria on \DeltaE, \mes, the candidate masses of the $\Xz$
and \KS, and the particle identification (PID) requirements of
leptons, pions, and charged kaons. The criteria $|\DeltaE|<15\mev$,
$|\mes-m_B|<6\mevcc$ and $|m(\jpsi\pip\pim)-3872\mevcc|<6\mevcc$
(signal region) were found to be optimal for selecting signal events.
In case of multiple candidates in an event, we select the candidate
with the smallest value of $|\DeltaE|$.  Applying our optimized
selection criteria, we compute the $\jpsi\pip\pim$ invariant mass in
the range $3.8-3.95\gevcc$ shown in Figs.~\ref{fig:fit}(a)
and~\ref{fig:fit}(b) for the $\Bm$ and $\Bz$ mode, respectively. The
shaded area shows events in the sideband region $|\mes-5260|<6\mevcc$.

We extract the number of signal events with an extended unbinned
maximum-likelihood fit to the two-dimensional distribution
$y(\mes,m_X)$ where $m_X$ is the $\jpsi\pip\pim$ invariant mass. The
probability density function (PDF) (normalized to the total number of
events) is $\mathcal{P}(y) = \sum_{t}n_t\mathcal{P}_t(y)$ where $n_t$
is the number of events of category $t$. We consider three different
event categories: signal, $B$ decays with the same final-state
particles as the signal that accumulate near $\mes\approx m_B$
(peaking background), and combinatorial background.  The individual
PDFs $\mathcal{P}_t$ are assumed to be uncorrelated in \mes and $m_X$
and can therefore be factorized as $\mathcal{P}_t(y) =
g_t(\mes)h_t(m_X)$, where $g_t$ and $h_t$ represent the $\mes$ and
$m_X$ probability distributions, respectively.  The $B\to\Xz K$ signal
events are modeled by a Gaussian distribution in \mes. The resolution
function in $m_X$ for those events is best described by a Cauchy
function~\cite{Cauchy} due to the mass constraint of the \jpsi candidate.
The PDF for peaking background events is
parameterized by a Gaussian distribution in \mes and a linear function
in $m_X$. We model combinatorial background events by an ARGUS
function~\cite{Albrecht:1986nr} in \mes and a linear function in
$m_X$. The fit performance was validated with MC experiments. The mean
and width of the \mes Gaussian distribution for signal and peaking
background and the width of the $m_X$ Cauchy distribution for the \Bz
mode were fixed to values obtained from MC samples. Other parameters
are allowed to vary in the fit.

The fit is performed in the region $5.2<\mes<5.3\gevcc$ and
$3.80<m_X<3.95\gevcc$ without applying the optimized selection criteria on
those two variables. The signal region projections of the
two-dimensional fit are shown in Fig.~\ref{fig:fit} for the $B^-$
(a,c) and $B^0$ (b,d) modes.
We obtain $61.2\pm15.3$ signal events for the $\Bm$ mode ($n_s^-$) and
$8.3\pm4.5$ signal events for the $\Bz$ mode ($n_s^0$), respectively.
In the following we interpret the observed events in the $\Bz$ mode as the
$X(3872)$.

The efficiency is determined from MC samples with an \Xz
signal of zero width at $3.872\gevcc$. The decay model consists of the
sequential isotropic decays $B\to\Xz K$, $\Xz\to\jpsi\rho^0$, and
$\rho^0\to\pip\pim$. Compared to a three-body decay, this gives a more
accurate description of the observed $\pip\pim$ invariant mass
distribution~\cite{Aubert:2004ns}.  Efficiencies are corrected for the
small differences in PID and tracking efficiencies that are found by
comparing data and MC control samples. The final efficiencies are
$(17.4\pm0.2)\%$ for the \Bz/\KS mode and $(22.2\pm0.2)\%$ for the \Bm/\Km
mode.

The branching fraction systematic errors ($\Bm$, $\Bz$ mode in \%)
include uncertainties in the number of \BB events (1.1, 1.1), secondary
branching fractions (5.0, 5.0)~\cite{pdg}, efficiency calculation due
to limited MC statistics (0.7, 1.9), MC decay model of the \Xz
(1.0, 1.6), differences between data and MC (1.8, 8.9), PID (5.0, 5.0),
charged particle tracking (6.0, 4.8), and \KS reconstruction
(-, 1.6). The production ratio of \Bz and \Bm mesons in \FourS decays
is $1.006\pm0.048$~\cite{Aubert:2004ur}. The total fractional error
obtained by adding the uncertainties in quadrature is $9.6\%$ and
$12.8\%$ for the $\Bm$ and $\Bz$ mode, respectively.

Assuming Gaussian systematic errors with a PDF 
$P_{sys}(n)\sim\exp[-(n-n_S)^2/2\sigma_{sys}^2]$, the negative log-likelihood (NLL)
function including systematic errors is
$L_{sys}=([1/L(n)]-[1/\ln P_{sys}(n)])^{-1}$ where
$L(n)=-\ln(\mathcal{L}(n)/\mathcal{L}_{max})$ is the NLL projection of
the parameter estimate $n$ of the number of signal events and
$\sigma_{sys}$ is the systematic error on the number of signal events.
The significance including systematic errors obtained from
$\sqrt{2L_{sys}(n=0)}$ is $2.5\sigma$ for the \Bz mode and $6.1\sigma$
for the \Bm mode. The statistical significance ($\sigma_{sys}=0$) of the
signal is $2.6\sigma$ and $7.5\sigma$, respectively.

Using $n^0_S$ and $n^-_S$, the efficiencies, the secondary branching
fractions and the number of \BB events, we obtain the branching
fractions
$\BR^0\equiv\BR(\Bz\to\Xz\Kz,X\to\jpsi\pip\pim)=(5.1\pm2.8\pm0.7)\times
10^{-6}$ and
$\BR^-\equiv\BR(\Bm\to\Xz\Km,X\to\jpsi\pip\pim)=(10.1\pm2.5\pm1.0)\times
10^{-6}$. For the ratio of branching fractions, $R\equiv\BR^0/\BR^-$,
where most of the systematic errors cancel, we obtain
$R=0.50\pm0.30\pm0.05$. We calculate a $90\%$ confidence level (C.L.)
likelihood interval~\cite{Cowan} $[n_l,n_h]$ for the number of signal
events in the \Bz mode by solving the equation
$2L_{sys}(n_{l,h})=[\mbox{erf}^{-1}(0.95)]^2$. With $n_l=2.2$ and
$n_h=16.9$ the $90\%$ C.L. interval on $\BR^0$ is $1.34\times
10^{-6}<\BR^0<10.3\times 10^{-6}$. Using the same strategy, the
confidence interval on the ratio of branching fractions becomes
$0.13<R<1.10$ at $90\%$ C.L.

We measure the mass of the \Xz in both modes in reference to the
precisely measured \psitwos mass~\cite{pdg}. We fit the
$\jpsi\pip\pim$ invariant mass in the \psitwos and \Xz region and
calculate $m_X =
m_{X,\text{fit}}-m_{\psitwos,\text{fit}}+m_{\psitwos}$. The result for 
the \Bz mode is $(3868.6\pm1.2\pm0.2)\mevcc$ and
$(3871.3\pm0.6\pm0.1)\mevcc$ for the
\Bm mode, where the first error is the statistical uncertainty on
$m_{X,\text{fit}}$ and the second is the uncertainty on
$m_{\psitwos,\text{fit}}$ and $m_{\psitwos}$~\cite{pdg}. The mass
difference of the \Xz produced in \Bz and \Bm decays is $\Delta
m=(2.7\pm1.3\pm0.2)\mevcc$. The full width at half maximum of the
$X$-mass distribution from the fit on data is $(6.7\pm2.7)\mevcc$,
which is consistent with the MC-determined value of
$(5.4\pm0.1)\mevcc$. From this we calculate the $90\%$ C.L. upper
limit on the natural width as $\Gamma<4.1\mevcc$.

\begin{figure}[ht]
\begin{center}
  \includegraphics[width=0.35\textwidth]{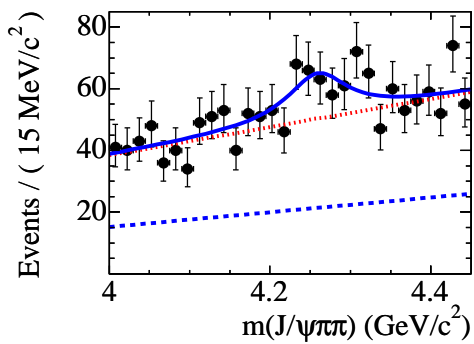}
  \caption{\label{fig:yzFit}
  Projection of the two-dimensional fit for the mass region above
  $4\gevcc$ for events within the \mes signal region in
  $\Bm\to\jpsi\pip\pim\Km$. The lines represent the same event types
as in Fig.~\ref{fig:fit}.}
\end{center}
\end{figure}

Recent observations by
\babar~\cite{Aubert:2005-ISR} in ISR events provide evidence for at least one
broad resonance in the invariant mass spectrum of $\jpsi\pip\pim$ at
$4.259\gevcc$ that can be characterized by a single resonance 
with a full width of $88\mevcc$. This structure is
referred to as $Y(4260)$. We search in $\Bm$ decays for states decaying into
$\jpsi\pip\pim$ above $4\gevcc$ and impose the additional selection
criterion $|m(\Km\pip\pim)-1273\mevcc|>250\mevcc$, which removes backgrounds
from $K_1(1270)$ decays. In the resulting mass distribution,
Fig.~\ref{fig:yzFit}, we observe large combinatoric backgrounds and
cannot reliably determine the parameters of one or more resonances.
We use a two-dimensional PDF identical to the previous model, but fix
the central value and width of the signal component to the ISR
results~\cite{Aubert:2005-ISR}. The natural width of $88\mevcc$ has
been enlarged by the detector resolution, which is found to be the same
as for the mass region around $3.87\gevcc$. The $m_X$ projection
of the two-dimensional fit is overlaid in Fig.~\ref{fig:yzFit} and yields
$128\pm42$ signal events. The statistical significance calculated from
$\sqrt{-2\ln\mathcal{L}_0/\mathcal{L}}$ is $3.1\sigma$ where
$\mathcal{L}$ and $\mathcal{L}_0$ are the maximum likelihood of the
fit and the null hypothesis fit, respectively. Using a phase-space MC
simulation of a state at $4.26\gevcc$ decaying into $\jpsi\pip\pim$
and assuming the same systematic uncertainties and efficiency
corrections as for the $X(3872)$, we obtain the $95\%$ C.L. upper limit
on the branching fraction
$\BR_Y=\BR(\Bm\to Y(4260)\Km,Y(4260)\to\jpsi\pip\pim)<2.9\times10^{-5}$.
The $90\%$ C.L. likelihood interval on the branching fraction
$\BR_Y=(2.0\pm0.7\pm0.2)\times10^{-5}$ is 
$1.2\times10^{-5}<\BR_Y<2.9\times10^{-5}$.

In conclusion, our studies of the $\jpsi\pip\pim$ invariant mass below $4\gevcc$
yield a signal of $2.5\sigma$ and $6.1\sigma$ significance in the $\Bz$ and
$\Bm$ mode, respectively, with a ratio of branching fractions
$R=0.50\pm0.30\pm0.05$. We observe an excess of events above background in the
$\jpsi\pip\pim$ invariant mass between $4.2$ and $4.4\gevcc$.
These events are consistent with the broad structure
observed in ISR events~\cite{Aubert:2005-ISR}.

If one narrow state
is observed in the mode $\Bm\to\Xz\Km$, $X\to\jpsi\pip\pim$, the
diquark-antidiquark model~\cite{Maiani:2004vq} predicts one amplitude
(from $X_{d}$ or $X_{u}$) to be dominant in the charged mode and the
other amplitude to be dominant in the neutral mode. In this case, the
model predicts the relative rates to be equal ($R=1$) and the mass
difference to be $(7\pm 2)\mevcc$. The ratio of branching fractions
is consistent with our measurement, $0.13<R<1.10$ at $90\%$ C.L., and the
observed mass difference of $(2.7\pm1.3\pm0.2)\mevcc$ is both
consistent with zero and the model prediction within two standard
deviations. In the S-wave molecule model~\cite{Braaten:2004ai}, the
neutral mode branching fraction is predicted to be at least 10 times
smaller ($R<0.1$) than the charged mode. However, we obtain a ratio of
neutral to charged branching fractions which is slightly more
consistent with isospin-conserving decays.

We are grateful for the excellent luminosity and machine conditions
provided by our \pep2\ colleagues, 
and for the substantial dedicated effort from
the computing organizations that support \babar.
The collaborating institutions wish to thank 
SLAC for its support and kind hospitality. 
This work is supported by
DOE
and NSF (USA),
NSERC (Canada),
IHEP (China),
CEA and
CNRS-IN2P3
(France),
BMBF and DFG
(Germany),
INFN (Italy),
FOM (The Netherlands),
NFR (Norway),
MIST (Russia), and
PPARC (United Kingdom). 
Individuals have received support from CONACyT (Mexico), A.~P.~Sloan Foundation, 
Research Corporation,
and Alexander von Humboldt Foundation.


\begin{thebibliography}{18}
\bibitem{chargeConj}
Charge conjugate reactions are included implicitly throughout.
\end{thebibliography}
\end{document}